\documentclass{article}[12pt]

\usepackage{amscd,amsmath,amsthm,amsfonts,amssymb}
\usepackage{graphicx}
\usepackage{multicol}
\usepackage{color}

\def\sech{{\rm sech}}

\def\PT{$\mathcal{PT}$}

\def\[{\begin{equation}}
\def\]{\end{equation}}

\begin{document}
\title{\bf Bifurcation of soliton families from linear modes in non-\PT-symmetric complex potentials}
\author{Sean Nixon and Jianke Yang \\
Department of Mathematics and Statistics \\
University of Vermont \\ Burlington, VT 05401, USA}
\date{}
\maketitle

\begin{abstract}
Continuous families of solitons in generalized nonlinear
Sch\"odinger equations with non-\PT-symmetric complex potentials are
studied analytically. Under a weak assumption, it is shown that
stationary equations for solitons admit a constant of motion if and
only if the complex potential is of a special form $g^2(x)+ig'(x)$,
where $g(x)$ is an arbitrary real function. Using this constant of
motion, the second-order complex soliton equation is reduced to a
new second-order real equation for the amplitude of the soliton.
From this real soliton equation, a novel perturbation
technique is employed to show that continuous families of solitons always
bifurcate out from linear discrete modes in these non-\PT-symmetric
complex potentials. All analytical results are corroborated by
numerical examples.
\end{abstract}

\section{Introduction}
Nonlinear wave systems fall into two major categories: conservative and
dissipative. Conservative systems are energy-conserving, and their
solitary waves (solitons) exist as continuous families
with continuous ranges of energy values. A typical example is the
nonlinear Schr\"odinger (NLS) equation. Dissipative systems contain
gain and loss, and their solitons are generally isolated with
certain discrete energy values. A typical example of this type is
the Ginzburg-Landau equation. A recent discovery is that, in
dissipative but parity-time (\PT) symmetric systems, solitons can
still exist as continuous families with continuous energy values
\cite{Bender1998,Musslimani2008, Wang2011, Lu2011, Abdullaev2011,
He2012, Nixon2012, Zezyulin2012a,Huang2013,Kartashov2013,Driben2011,
Alexeeva2012,Moreira2012,Konotop2012, Kevrekidis2013,Li2011,
Zezyulin2012b, Zezyulin2013,Wimmer2015}. An example in this category
is the NLS equation with a complex but \PT-symmetric potential.
These soliton families are allowed since the \PT symmetry assures
that the gain and loss of the soliton is perfectly balanced at
arbitrary energy levels.

In dissipative and non-\PT-symmetric systems, the expectation is
that any solitons will be isolated with discrete energy values, as seen in
typical dissipative systems \cite{Yang_PLA}. However, exceptions
were reported numerically in \cite{Tsoy2014,Konotop_OL2014} for the
NLS equation with a non-\PT-symmetric complex potential of special
form, where families of solitons with continuous energy values can
bifurcate out from the linear modes of the potential. This finding is
very surprising in view of the lack of \PT symmetry here. For these
special potentials, a constant of motion was discovered in
\cite{Konotop_OL2014} for the stationary soliton equation. Using
this constant of motion, soliton families in these special
potentials were explained by a numerical shooting argument
\cite{Konotop_OL2014}.

In this article, we analytically investigate solitons of the NLS
equation with non-\PT-symmetric complex potentials. We focus on
three main questions: (1) What types of non-\PT-symmetric complex
potentials admit soliton families? (2) How can one analytically
explain and calculate soliton families bifurcating from linear modes
in such potentials? (3) Do these soliton families exist under other
nonlinearities?

Regarding the first question, we recognize that in the absence of
\PT symmetry, the existence of a constant of motion in the
stationary soliton equation plays a crucial role in the existence
of soliton families. Assuming this constant of motion for complex
potentials is a continuous deformation of one that exists in the NLS equation without a potential, 
we show that the only complex potentials which admit a
constant of motion are those of the form reported in
\cite{Tsoy2014,Konotop_OL2014}, i.e., $V(x)=g^2(x)+ig'(x)$, where
$g(x)$ is an arbitrary real function. This strongly suggests that
potentials of the above form are the \emph{only} one-dimensional
non-\PT-symmetric complex potentials that admit soliton families.

On the second question, through use of the constant of motion, we
reduce the second-order complex soliton equation to a new
second-order real equation for the square of amplitude of the soliton, which
is then solved perturbatively for a continuous range of $\mu$ values.
This way, the existence of soliton families bifurcating from linear
modes in non-\PT-symmetric potentials is analytically explained and
explicitly calculated. Interestingly, this perturbation calculation
of solitons differs significantly from the method used for real and \PT-symmetric potentials
because the linearization operator of the new real equation has a
distinctly different kernel structure.

Regarding the third question, we show that these soliton families
still exist under a more general class of nonlinearities. Furthermore, the choice of nonlinearity within this class has no effect on the existence of a constant of motion.

These analytical results are compared with numerical
examples, and good agreement between them is illustrated.

\section{Preliminaries}

The mathematical model we consider, in most parts of this article,
is the NLS equation with a complex potential,
\[ \label{e:NLS}
i\Psi_t+\Psi_{xx}+V(x)\Psi+\sigma|\Psi|^2\Psi=0,
\]
where $\sigma=\pm 1$ is the sign of cubic nonlinearity. This model
describes paraxial nonlinear light propagation in a waveguide with
gain and loss \cite{Musslimani2008,Yang_SIAM}, as well as
Bose-Einstein condensates with atoms injected into one part of the
potential and removed from another part of the potential
\cite{Pitaevskii2003,BECPT}. Most of the earlier work focused on the
case where the complex potential $V(x)$ is \PT-symmetric, i.e.,
$V^*(x)=V(-x)$, with the superscript `*' representing complex
conjugation \cite{Musslimani2008, Wang2011, Lu2011, He2012,
Nixon2012, Zezyulin2012a,Huang2013}. In this article, we consider
the case where $V(x)$ is not \PT-symmetric, i.e.,
\[
V^*(x)\ne V(-x).
\]

Soliton solutions of equation \eqref{e:NLS} take the form
\[ \label{e:soliton}
\Psi(x, t)=\psi(x) e^{i\mu t},
\]
where $\psi(x)$ is a localized function solving the stationary
equation
\[ \label{e:psi}
\psi_{xx}-\mu\psi+V(x)\psi+\sigma|\psi|^2\psi=0,
\]
and $\mu$ is a real propagation constant. In \PT-symmetric
potentials, solitons exist as continuous families parameterized by
$\mu$ \cite{Wang2011, Lu2011, He2012, Nixon2012,
Zezyulin2012a,Huang2013}. But in non-\PT-symmetric potentials,
soliton families are generically forbidden \cite{Yang_PLA}.
Surprisingly, it was reported recently through numerical examples
that in complex potentials of the special form
\[ \label{e:Vx}
V(x)=g^2(x)+ig'(x),
\]
where $g(x)$ is an arbitrary real function, soliton families can
still bifurcate out from linear modes even when $V(x)$ is
non-\PT-symmetric (i.e., when $g(x)$ is not even)
\cite{Tsoy2014,Konotop_OL2014}. This result is very unintuitive.
Indeed, if one performs a regular perturbation calculation of soliton
families bifurcating from linear modes in a general complex
potential, it will be seen that infinitely many nontrivial
conditions would have to be satisfied simultaneously, which makes
such bifurcation almost impossible \cite{Yang_PLA}. However, for the
special complex potential (\ref{e:Vx}), all those conditions are
met, which is miraculous. Obviously this phenomenon needs better
understanding. A step in this direction was made in
\cite{Konotop_OL2014}, where through the discovery of a constant of
motion for the soliton equation (\ref{e:psi}) under the potential
(\ref{e:Vx}), soliton families in Eq. (\ref{e:psi}) were explained
through a numerical shooting argument.

Many important questions are currently open regarding soliton
families in non-\PT-symmetric complex potentials. For instance, what
other non-\PT-symmetric complex potentials admit soliton families?
How can one analytically explain and explicitly calculate soliton
families bifurcating from linear modes in non-\PT-symmetric
potentials? Do these soliton families also exist under other
nonlinearities? These questions will be investigated in the
remainder of this article.

\section{Constant of motion} \label{sec:3}

A quantity $J(x, \psi)$ is called a constant of motion in the
stationary equation (\ref{e:psi}) if $dJ/dx=0$. The existence of a
constant of motion proves to be important for the existence of
soliton families (see \cite{Konotop_OL2014} and later text). Thus, in
this section, we study what complex potentials $V(x)$ admit a
constant of motion. In this study, solutions, $\psi(x)$, to the
stationary equation (\ref{e:psi}) are allowed to be any solutions,
not necessarily solitons. That is, $\psi(x)$ is allowed to be
non-local.

First we split the complex potential $V(x)$ into real and imaginary
parts,
\[
V(x)=v_1(x)+iv_2(x),
\]
where $v_1(x), v_2(x)$ are real functions. We also express the
complex function $\psi(x)$ in polar forms,
\[  \label{e:psipolar}
\psi(x)=r(x) e^{i\int \theta(x) dx},
\]
where $r(x)$, $\theta(x)$ are real amplitude and phase functions.
Substituting these expressions into the soliton equation
(\ref{e:psi}), we get
\begin{eqnarray}
r_{xx}-\mu r+v_1r+\sigma r^3-\theta^2r =0,  \label{e:rxx}   \\
(r^2\theta)_x = -v_2r^2.                    \label{e:r2theta}
\end{eqnarray}

In the absence of the potential ($v_1=v_2=0$), it is easy to verify that this system admits two constants of motion
\[
J_1=r^2\theta,
\]
and
\[
J_2=r_x^2-\mu r^2 +\frac{\sigma}{2}r^4+r^2\theta^2,
\]
where $dJ_1/dx=dJ_2/dx=0$. Since the system is third order, these
are the only constants of motion the system can allow. These two
constants of motion are associated with the flux terms of the power
and momentum conservation laws of the potential-free NLS equation,
but this fact is not important to our analysis.

In the presence of the potential, it is reasonable to assume that
the corresponding constant of motion $J$ is a continuous deformation
of those in the potential-free case. In other words, $J$ approaches
constants of motion of the potential-free equation when $v_1, v_2$
approach zero. Notice that $J_1$ and $J_2$ have different ranks
\cite{Hereman}. Thus under the limit $v_1, v_2 \to 0$, $J$ can only
approach one of $(J_1, J_2)$, not their linear combination. Our
strategy then is to calculate $dJ_k/dx \: (k=1, 2)$ in the presence
of the potential and derive conditions on $(v_1, v_2)$ so that
$dJ_k/dx$ is a total derivative of $x$, i.e., a constant of motion
is admitted.

Now we calculate $dJ_k/dx$ in the presence of a potential. First
we consider $dJ_1/dx$. Eq. (\ref{e:r2theta}) clearly shows that, in
order for $dJ_1/dx$ to be a total derivative, we must have $v_2=0$,
i.e., the potential $V(x)$ is real. This is not what we want since
we exclusively consider complex potentials in this paper. Thus, there
are no constants of motion in Eqs. (\ref{e:rxx})-(\ref{e:r2theta})
that approach $J_1$ when the complex potential approaches zero.

Next we consider $dJ_2/dx$. Utilizing equations
(\ref{e:rxx})-(\ref{e:r2theta}), we readily find that
\[
\frac{dJ_2}{dx}=-v_1(r^2)_x-2v_2 r^2\theta.
\]
The right side of this equation can be rewritten as
\[
\frac{dJ_2}{dx}=W_x+r^2v_{1x}+2(r^2\theta)_x\int v_2 dx,
\]
where
\[
W=-v_1r^2-2r^2\theta \int v_2 dx.  \nonumber
\]
Then utilizing equation (\ref{e:r2theta}), the above equation
becomes
\[
\frac{dJ_2}{dx}=W_x+r^2\left(v_{1x}-2v_2\int v_2 dx\right).
\]
In order for the right side of the above equation to be a total
derivative, the necessary and sufficient condition is
\[
v_{1x}=2v_2\int v_2 dx.
\]
This condition can be rewritten as
\[
v_{1x}=\left[\left(\int v_2 dx\right)^2\right]_x,
\]
thus
\[
v_1=\left(\int v_2 dx\right)^2+C,
\]
where $C$ is an arbitrary constant. Finally, denoting
\[
g=\int v_2 dx,
\]
the potential $V(x)$ which admits a constant of motion then is of
the form
\[
V(x)=g^2(x)+ig'(x)+C.
\]
Obviously, the constant $C$ in this potential can be eliminated from
Eq. (\ref{e:NLS}) through a simple gauge transformation. The
remaining potential is then of the form (\ref{e:Vx}). Thus we
conclude that if the constant of motion for the stationary equation
(\ref{e:psi}) with a complex potential is a continuous deformation
of $J_2$ without the potential, then this constant of motion exists
if and only if the complex potential $V(x)$ is of the special form
(\ref{e:Vx}), and the corresponding motion constant is
\[
J=J_2-W=J_2+g^2r^2+2gr^2\theta,
\]
or more explicitly,
\[  \label{e:Cmotion}
J=r_x^2-\mu r^2 +\frac{\sigma}{2}r^4+r^2(\theta+g)^2,
\]
where $dJ/dx=0$. This constant of motion agrees with that reported
in \cite{Konotop_OL2014} for these special potentials (\ref{e:Vx}).

\section{Bifurcation of soliton families}

In this section, we analytically calculate the bifurcation of
solitons from linear modes in Eq. (\ref{e:NLS}), with potential of the special form (\ref{e:Vx}), and show that soliton families bifurcate out
in such non-\PT-symmetric systems.

For the potential (\ref{e:Vx}), when solitons (\ref{e:soliton}) are
expressed in polar forms (\ref{e:psipolar}), the equations for $r$
and $\theta$ are seen from Eqs. (\ref{e:rxx})-(\ref{e:r2theta}) as
\begin{eqnarray}
r_{xx}-\mu r+g^2r+\sigma r^3-\theta^2r =0,  \label{e:r} \\
(r^2\theta)_x = -g_xr^2,
\end{eqnarray}
and these equations admit a constant of motion (\ref{e:Cmotion}).
For solitons, this constant $J$ can be evaluated at $x=\infty$ as
zero, thus
\[  \label{e:C}
r_x^2-\mu r^2 +\frac{\sigma}{2}r^4+r^2(\theta+g)^2=0.
\]
From this equation, we get
\[
\theta=-g\pm \sqrt{\mu-\frac{1}{2}\sigma r^2-\frac{r_x^2}{r^2}}.
\]
Inserting it into Eq. (\ref{e:r}) and after simple algebra, we get
\[
r_{xx}-2\mu r+\frac{3}{2}\sigma r^3+\frac{r_x^2}{r}=\mp 2g\sqrt{\mu r^2-\frac{1}{2}\sigma r^4-r_x^2},
\]
or
\[
(r^2)_{xx}-4\mu r^2+3\sigma r^4\pm 2g \sqrt{4\mu r^4-2\sigma r^6-[(r^2)_x]^2}=0.
\]
Denoting $R=r^2$, we arrive at a single second-order equation for
the real amplitude function $R$ as
\[  \label{e:Rpm}
R_{xx}-4\mu R+3\sigma R^2\pm 2g \sqrt{4\mu R^2-2\sigma R^3-R_x^2}=0,
\]
which can also be rewritten as
\[ \label{e:Rform2}
\left(\sqrt{4\mu R^2-2\sigma R^3-R_x^2}\right)_x=\pm 2gR_x.
\]

\subsection{Perturbation calculations}

The sign in Eq. (\ref{e:Rpm}) needs to be chosen appropriately
according to the function $g(x)$. Indeed, if $g(x)$ switches to
$-g(x)$, this sign should switch as well. Without loss of
generality, we take the plus sign in Eq. (\ref{e:Rpm}),
\[ \label{e:R}
R_{xx}-4\mu R+3\sigma R^2+ 2g \sqrt{4\mu R^2-2\sigma R^3-R_x^2}=0.
\]
Note that sometimes the same solution $R(x)$ can lead to mixed signs
in Eq. (\ref{e:Rpm}) on different $x$-intervals. This could occur if
$4\mu R^2-2\sigma R^3-R_x^2$ is zero somewhere on the $x$-axis,
since the square root is a possible mechanism for inducing a sign
change so that the square-rooted quantity remains smooth. We do not
consider such mixed cases here. This exclusion will be assured by
Assumption~1 in Sec.~\ref{subsec:solvability}.

For a localized function $g(x)$, it is easy to see that the
large-$x$ asymptotics of the soliton solution in Eq. (\ref{e:R})
are, to leading order,
\[ \label{e:Rasym}
R(x) \to a_{\pm} e^{-\sqrt{4\mu}|x|}, \quad x\to \pm \infty,
\]
where $a_{\pm}$ are positive constants.

First we consider linear modes in Eq. (\ref{e:R}), which satisfy the
equation
\[ \label{e:phi}
\phi_{xx}-4\mu_0\phi+2g\sqrt{4\mu_0\phi^2-\phi_x^2}=0.
\]
This equation can be rewritten as
\[ \label{e:phiform2}
\left(\sqrt{4\mu_0 \phi^2-\phi_x^2}\right)_x=2g\phi_x.
\]
Since $g$ and $\phi$ are localized functions, we see that
\[ \label{e:sqrtformula}
\sqrt{4\mu_0 \phi^2-\phi_x^2}=\int_{-\infty}^x 2g\phi_\xi d\xi=
-\int_x^\infty 2g\phi_\xi d\xi,
\]
and
\[
\int_{-\infty}^\infty 2g\phi_\xi d\xi=0.
\]

Equation (\ref{e:phi}) is scaling-invariant and thus an eigenvalue
problem, but it is nonlinear in both the eigenvalue $\mu_0$ and
eigenfunction $\phi$. Thus, this is a different type of eigenvalue
problem. Solving this new eigenvalue problem is equivalent to
solving for discrete real eigenmodes in the original eigenvalue
problem from Eq.~(\ref{e:psi}), i.e.,
\[ \label{e:psiVx}
\psi_{xx}+V(x)\psi=\mu_0\psi,
\]
and the eigenfunction correspondence is $\phi=|\psi|^2$. Previous
results in \cite{Tsoy2014} have shown that for the underlying
special potential (\ref{e:Vx}), the linear eigenvalue problem
(\ref{e:psiVx}) admits discrete real eigenvalues for a large class
of functions $g(x)$. The new eigenvalue problem (\ref{e:phi}) makes
the existence of such real eigenvalues more clear since all
quantities in that equation are real.

From such eigenmodes, families of solitons can bifurcate out under
variation of $\mu$. We will analytically prove this by explicitly
calculating this soliton bifurcation from a linear mode $(\mu_0,
\phi)$ using perturbation methods.

The perturbation expansion is
\begin{eqnarray}
R&=&\epsilon (R_0+\epsilon R_1+\epsilon^2R_2+\dots),   \label{e:Rseries}\\
\mu &=& \mu_0+\epsilon,       \label{e:museries}
\end{eqnarray}
where $\epsilon >0$ is a small parameter. Here we have assumed the
bifurcation occurs to the right side of $\mu_0$. As we will see
later, this assumption dictates the sign of nonlinearity $\sigma$.
If the bifurcation occurs to the left side of $\mu_0$, then only
trivial modifications to our analysis are needed, and the
bifurcation will occur for the opposite sign of nonlinearity.

Inserting the above expansion into Eq. (\ref{e:R}), at order
$\epsilon$, we find
\[  \label{e:R0}
R_0=c_0\phi,
\]
where $c_0$ is a positive constant to be determined.

At order $\epsilon^2$, we get
\[  \label{e:R1}
LR_1=F,
\]
where
\[
L=\partial_{xx}+p_1\partial_x+p_2,
\]
\[ \label{e:p1p2}
p_1=-\frac{2g\phi_x}{\sqrt{4\mu_0\phi^2-\phi_x^2}}, \quad p_2=4\mu_0\left(\frac{2g\phi}{\sqrt{4\mu_0\phi^2-\phi_x^2}}-1\right),
\]
\[ \label{e:Fformula}
F=c_0(f_1-c_0\sigma f_2),
\]
and
\[
f_1=4\phi\left(1-\frac{g\phi}{\sqrt{4\mu_0\phi^2-\phi_x^2}}\right), \quad f_2=\phi^2\left(3-\frac{2g\phi}{\sqrt{4\mu_0\phi^2-\phi_x^2}}\right).
\]

Now it is time to analyze the properties of homogeneous solutions
and adjoint homogeneous solutions of the operator $L$ and the
solvability condition of Eq.~(\ref{e:R1}).

\subsection{Kernels of linearization operators $L$ and $L^A$}

First, it is easy to verify that $\phi$ is a homogeneous solution of
$L$, i.e.,
\[  \label{e:Lphi}
L\phi=0.
\]
Let us suppose the other homogeneous solution of $L$ is $\phi_2$,
then according to Abel's formula, the Wronskian of $(\phi, \phi_2)$
is
\[
W(\phi, \phi_2)=W_0 e^{-\int p_1 dx},
\]
or
\[ \label{e:W1}
W(\phi, \phi_2)=W_0\sqrt{4\mu_0\phi^2-\phi_x^2}
\]
in view of Eqs. (\ref{e:phiform2}) and (\ref{e:p1p2}). Here $W_0$ is
a constant. Utilizing Eq. (\ref{e:sqrtformula}), the above Wronskian
can be rewritten as
\[ \label{e:W}
W(\phi, \phi_2)=W_0\int_{-\infty}^x 2g\phi_\xi d\xi=-W_0
\int_x^\infty 2g\phi_\xi d\xi.
\]
From this formula we see that, if $g(x)$ is a localized function,
then the decay rate of this Wronskian at large $|x|$ is faster than
that of $\phi$, and thus $\phi_2$ is also a localized function.

Using these homogeneous solutions of $L$, we can build homogeneous
solutions of the adjoint operator $L^A$, where
\[
L^A=\partial_{xx}-\partial_xp_1+p_2.
\]

\textbf{Lemma 1} The two homogeneous solutions $(\phi^A, \phi_2^A)$
of adjoint operator $L^A$ are
\[  \label{e:phiA2A}
\phi^A=-\frac{\phi_2}{W(\phi, \phi_2)}, \quad
\phi_2^A=\frac{\phi}{W(\phi, \phi_2)}.
\]

Proof: We first turn the second-order homogeneous equation of
operator $L$ into a system of first-order equations,
\[ \label{e:Xprime}
X'=\left[\begin{array}{cc} 0 & 1 \\ -p_2 & -p_1\end{array}\right]X,
\]
where the prime stands for derivative to $x$. The fundamental matrix
solution to this system is
\[
X=\left[\begin{array}{cc} \phi & \phi_2 \\ \phi_x & \phi_{2x} \end{array}\right].
\]
The adjoint system of Eq. (\ref{e:Xprime}) is
\[  \label{e:Yprime}
Y'=-\left[\begin{array}{cc} 0 & -p_2 \\ 1 & -p_1\end{array}\right]Y.
\]
Notice that if $Y=[y_1, y_2]^T$, where the superscript `$T$'
represents vector or matrix transpose, then it is easy to verify
that
\[
L^A y_2=0,
\]
i.e., the second component of vector solution $Y$ is in the kernel
of the adjoint operator $L^A$.

It is well known that the fundamental matrix solution to the adjoint
vector system (\ref{e:Yprime}) is $\left(X^{-1}\right)^T$. This can
be proved by calculating $(XX^{-1})'$, where upon utilizing Eq.
(\ref{e:Xprime}), a homogeneous differential equation for $X^{-1}$
would be obtained. Taking the transpose of this equation would
reveal that $\left(X^{-1}\right)^T$ satisfies the adjoint equation
(\ref{e:Yprime}). Notice that
\[
\left(X^{-1}\right)^T=\frac{1}{W(\phi, \phi_2)} \left[\begin{array}{cc} \phi_{2x} & -\phi_x \\ -\phi_2 & \phi \end{array}\right].
\]
Since the second-row functions in this matrix are in the kernel of
the adjoint operator $L^A$, the functions $\phi^A$ and $\phi_2^A$
defined in Eq. (\ref{e:phiA2A}) are then homogeneous solutions of
the adjoint operator $L^A$.   $\Box$

In view of Lemma 1, if $g(x)$ is a localized function, then both
adjoint homogeneous solutions $\phi^A, \phi_2^A$ are unbounded,
because the decay rates of $\phi$ and $\phi_2$ at large $|x|$ are
slower than that of the Wronskian $W(\phi, \phi_2)$.

The fact that $L$ has only localized solutions and $L^A$ has
only unbounded solutions in their kernels makes the solvability
condition for the first-order equation (\ref{e:R1}) novel, as we
will delineate below.

\subsection{Solvability conditions for certain potentials}  \label{subsec:solvability}

In this subsection, we show how to impose the solvability condition
on Eq. (\ref{e:R1}) under the following assumptions.

\vspace{0.3cm} \textbf{Assumption 1 } \ For the linear eigenmode
$(\mu_0, \phi)$, $4\mu_0\phi^2-\phi_x^2$ is strictly positive for
all $x$;

\textbf{Assumption 2 } \ The function $g(x)$ decays exponentially at
large $x$ as
\[
g(x) \to b_\pm e^{-\beta |x|}, \quad x \to \pm \infty,
\]
where $b_\pm$ and $\beta>0$ are constants.

\textbf{Assumption 3 } For these potentials, $\sqrt{4\mu_0} >
\beta$.

Assumption 1 assures that the linear operators $L, L^A$ are
nonsingular. In addition, there will be no sign change on the
$x$-interval in Eq. (\ref{e:Rpm}). This assumption will be made
throughout the text.

Assumptions 2 and 3 are introduced in order to make our analysis
more explicit. If the function $g(x)$ does not satisfy these
assumptions, an alternative analysis will be outlined in the next
subsection.

\textbf{Remark 1 }\, In some sense Assumptions 2 and 3 represent the
most common case, since for practical purposes the exact decay rate
at large $x$ should have minimal effect on the dynamics of the
system. Hence modifying the small tails of the potential to have a
suitably exponentially decaying rate should not make a meaningful
difference. However, we will still show in the next subsection that
these perturbation calculations may be performed for general
potentials, and then verify all results numerically.

At large $|x|$, the asymptotics of the eigenfunction $\phi(x)$ can
be readily seen from Eq. (\ref{e:phi}) as
\[ \label{e:phiasym}
\phi(x) \to \gamma_\pm e^{-\sqrt{4\mu_0} |x|}, \quad x\to \pm \infty,
\]
where $\gamma_\pm$ are constants. Then under Assumption 2, it is
easy to see from Eqs. (\ref{e:sqrtformula}) that the large-$x$
asymptotics of $\sqrt{4\mu_0\phi^2-\phi_x^2}$ is
\[  \label{e:asymsqrt}
\sqrt{4\mu_0\phi^2-\phi_x^2} \to s_\pm e^{-(\beta+\sqrt{4\mu_0})|x|},
\quad x \to \pm \infty,
\]
where
$$
s_\pm =\frac{2b_\pm\gamma_\pm\sqrt{4\mu_0}}{\beta+\sqrt{4\mu_0}}.
$$
Thus
\begin{eqnarray}
& p_1(x) \to  \pm (\beta+\sqrt{4\mu_0}), \quad & x \to \pm \infty, \label{e:asymp1} \\
& p_2(x) \to  \beta\sqrt{4\mu_0},\quad & x \to \pm \infty,
\end{eqnarray}
hence the asymptotics of operators $L$ and $L^A$ are
\[
L \to \partial_{xx}\pm (\beta+\sqrt{4\mu_0})\partial_x+\beta\sqrt{4\mu_0}, \quad x \to \pm \infty,
\]
and
\[
L^A \to \partial_{xx}\mp (\beta+\sqrt{4\mu_0})\partial_x+\beta\sqrt{4\mu_0}, \quad x \to \pm \infty.
\]
From these asymptotics, it is seen more explicitly that all
homogeneous solutions of $L$ are localized (as $e^{-\beta |x|}$,
$e^{-\sqrt{4\mu_0}|x|}$, or their linear combinations), and all
homogeneous solutions of $L^A$ are unbounded (as $e^{\beta |x|}$,
$e^{\sqrt{4\mu_0}|x|}$, or their linear combinations).

Regarding the second homogeneous solution $\phi_2(x)$, in view of
the asymptotics (\ref{e:phiasym}) of the first homogeneous solution
$\phi(x)$, without loss of generality we can set the large
negative-$x$ asymptotics of $\phi_2(x)$ as
\[ \label{e:asymphi2l}
\phi_2(x) \to e^{\beta x}, \quad x \to -\infty.
\]
Then its large positive-$x$ asymptotics is
\[ \label{e:asymphi2r}
\phi_2(x) \to \kappa_1 e^{-\beta x} +\kappa_2e^{-\sqrt{4\mu_0}x} , \quad x \to +\infty,
\]
where $\kappa_1, \kappa_2$ are constants. Substituting these
asymptotics into the Wronskian function and using the Wronskian
formula (\ref{e:W}), the value of $\kappa_1$ can be determined. But
this $\kappa_1$ value is not needed in our analysis.

From Lemma 1 and Eq. (\ref{e:W1}), we rewrite the adjoint
homogeneous solutions $\phi^A$ and $\phi_2^A$ equivalently as
\[
\phi^A=-\frac{\phi_2}{\sqrt{4\mu_0\phi^2-\phi_x^2}}, \quad
\phi_2^A=\frac{\phi}{\sqrt{4\mu_0\phi^2-\phi_x^2}}.
\]
Then using the asymptotics (\ref{e:phiasym}), (\ref{e:asymsqrt}),
(\ref{e:asymphi2l}) and (\ref{e:asymphi2r}), we find that the
large-$x$ asymptotics of $\phi^A$ and $\phi_2^A$ are
\[
\phi^A(x) \to \left\{ \begin{array}{ll} -d_{-}e^{-\sqrt{4\mu_0}x}, \quad & x \to -\infty, \\
-d_{+}\left(\kappa_1 e^{\sqrt{4\mu_0}x}+\kappa_2 e^{\beta x}\right), \quad & x\to +\infty,
\end{array}\right.
\]
and
\[ \label{e:asymphi2A}
\phi_2^A(x) \to j_\pm e^{\beta |x|}, \quad x \to \pm \infty,
\]
where
\[ \label{f:djpm}
d_{\pm}=\frac{\beta+\sqrt{4\mu_0}}{2b_\pm\gamma_\pm\sqrt{4\mu_0}}, \quad j_\pm=d_\pm \gamma_\pm.
\]
Notice that both adjoint solutions are unbounded and grow
exponentially at large $x$.

The asymptotics of functions $f_1$ and $f_2$ in the first-order
equation (\ref{e:R1}) can be similarly obtained as
\[ \label{e:asymf1}
f_1(x) \to q_\pm e^{-\sqrt{4\mu_0}|x|}, \quad x \to \pm \infty,
\]
and
\[ \label{e:asymf2}
f_2(x) \to w_\pm e^{-2\sqrt{4\mu_0}|x|}, \quad x \to \pm \infty,
\]
where
\[
q_\pm =2\left(1-\frac{\beta}{\sqrt{4\mu_0}}\right)\gamma_\pm, \quad
w_\pm =\left(2-\frac{\beta}{\sqrt{4\mu_0}}\right)\gamma_\pm^2.
\]

Now we consider the solvability condition of the first-order
equation (\ref{e:R1}). We see from Eqs. (\ref{e:Rasym}),
(\ref{e:R1}), (\ref{e:Fformula}) and (\ref{e:asymf2}) that the
large-$x$ asymptotics of $R_1(x)$ must be
\[ \label{e:R1asym}
R_1(x) \to P_1^\pm (x) e^{-\sqrt{4\mu_0}|x|} + C_1^\pm e^{-2\sqrt{4\mu_0} |x|}, \quad x \to \pm \infty,
\]
where $P_1^\pm(x)$ are certain linear functions of $x$ (these linear
functions come about when one expands the tail function
$e^{-\sqrt{4\mu}|x|}$ of (\ref{e:Rasym}) into a perturbation series
around $\mu=\mu_0$), and $C_1^\pm$ are constants. Note that the
tails $e^{-2\sqrt{4\mu_0} |x|}$ in the above equation are induced by
the nonlinearity-related forcing term $f_2$ and are admissible. They
do not contradict the leading-order $R(x)$ asymptotics
(\ref{e:Rasym}) since they are of higher order. Enforcement of this
tail behavior for $R_1(x)$ will yield the solvability condition
which determines the $c_0$ value.

We begin by taking the inner product of Eq. (\ref{e:R1}) with
$\phi_2^A(x)$ to get
\[ \label{e:solva1}
\langle \phi_2^A, LR_1\rangle = \langle \phi_2^A,  c_0(f_1-c_0\sigma f_2) \rangle,
\]
where the inner product is defined as
\[
\langle f, g\rangle \equiv \int_{-\infty}^\infty f^* g dx.
\]
Performing integration by parts, the left side of this equation
becomes
\begin{eqnarray}
\langle \phi_2^A, LR_1\rangle & = & \langle L^A\phi_2^A, R_1\rangle +\left.\left(\phi_2^AR_{1x}-\phi_{2x}^AR_1+p_1\phi_2^AR_1\right)\right|_{-\infty}^{+\infty}  \nonumber
\\ &=& \left.\left(\phi_2^AR_{1x}-\phi_{2x}^AR_1+p_1\phi_2^AR_1\right)\right|_{-\infty}^{+\infty}.    \label{e:bc}
\end{eqnarray}
In view of the asymptotics of $\phi_2^A, R_1$ and $p_1$ in Eqs.
(\ref{e:asymp1}), (\ref{e:asymphi2A}) and (\ref{e:R1asym}), as well
as Assumption 3, we see that the right side of the above equation is
zero, hence we obtain a solvability condition from Eq.
(\ref{e:solva1}) as
\[
\langle \phi_2^A,  f_1-c_0\sigma f_2\rangle=0.
\]
This solvability condition is the analog of Fredholm Alternatives
condition, and it quickly yields the formula for $c_0$ as
\[  \label{e:c0}
c_0=\frac{\langle \phi_2^A,  f_1\rangle}{\sigma\langle \phi_2^A,  f_2\rangle}.
\]
Notice that $f_1$ and $f_2$ decay at large $x$ as
$e^{-\sqrt{4\mu_0}|x|}$ or faster [see Eqs.
(\ref{e:asymf1})-(\ref{e:asymf2})], and $\phi_2^A(x)$ grows at large
$x$ as $e^{\beta |x|}$ [see (\ref{e:asymphi2A})]. Thus under
Assumption 3, both integrals in the inner products of the above
equation converge, and hence $c_0$ is well defined.

Eq. (\ref{e:c0}) is a necessary condition for the existence of the
first-order solution $R_1(x)$ with suitable asymptotics
(\ref{e:R1asym}). Since $c_0$ must be positive, Eq. (\ref{e:c0})
then shows that, in order for the soliton bifurcation to occur to
the right side of $\mu_0$ [see Eq. (\ref{e:museries})], the sign of
nonlinearity $\sigma$ must be chosen as the sign of the ratio
$\langle \phi_2^A, f_1\rangle/\langle \phi_2^A, f_2\rangle$.

The above solvability condition (\ref{e:c0}) turns out to be also
sufficient for the existence of solution $R_1(x)$ with suitable
asymptotics (\ref{e:R1asym}). To show this, we notice that the
general solution to the first-order equation (\ref{e:R1}) can be
derived by variation of parameters as
\[ \label{f:R1}
R_1(x)=\phi(x)\int_0^x \phi^A(\xi)F(\xi)d\xi + \phi_2(x) \int_0^x \phi_2^A(\xi)F(\xi)d\xi
+c_1\phi(x)+c_2\phi_2(x),
\]
where $c_1$ and $c_2$ are real constants. Using the asymptotics
detailed earlier in this section and under Assumption 3, we find
that at large $x$, $\phi^A(x)F(x)$ approaches a constant, and
$\phi_2^A(x)F(x)$ decays exponentially. Thus
\[
\int_0^x \phi^A(\xi)F(\xi)d\xi \to  \widetilde{P}_1^\pm(x),  \quad x\to \pm\infty,
\]
where $\widetilde{P}_1^\pm(x)$ are linear functions of $x$, and
\[
\int_0^x \phi_2^A(\xi)F(\xi)d\xi \to  \int_0^{\pm\infty} \phi_2^A(\xi)F(\xi)d\xi,  \quad x\to \pm\infty.
\]
In view of the large-$x$ asymptotics of $\phi(x)$ and $\phi_2(x)$,
in order for $R_1(x)$ in (\ref{f:R1}) to exhibit the suitable
asymptotics (\ref{e:R1asym}), the necessary and sufficient
conditions are
\[ \label{e:c2}
\int_0^{\pm\infty} \phi_2^A(\xi)F(\xi)d\xi+c_2=0,
\]
which leads to the equation
\[
\int_{-\infty}^{+\infty} \phi_2^A(x)F(x)dx=0.
\]
Substituting the expression (\ref{e:Fformula}) for $F(x)$ into this
equation, we then obtain the $c_0$ formula (\ref{e:c0}). Hence this
$c_0$ formula is a necessary and sufficient condition for the
existence of solution $R_1(x)$ with suitable asymptotics
(\ref{e:R1asym}).

In the $R_1$ formula (\ref{f:R1}), while $c_0$ is given by formula
(\ref{e:c0}) and $c_2$ given by equation (\ref{e:c2}), $c_1$ is
still a free parameter. This $c_1$ parameter will be fixed by
requiring the second-order solution $R_2(x)$ to have suitable
large-$x$ asymptotics [similar to (\ref{e:R1asym}) but with linear
functions $P_1^\pm(x)$ replaced by quadratic functions
$P_2^\pm(x)$]. This calculation of $c_1$ is in the same spirit of
the $c_0$ calculation, thus details will not be pursued in this
article.

\subsection{Extension to general potentials}  \label{subsec:Generalization}

In the event that Assumptions 2 and 3 of the previous subsection do
not hold, i.e., the decay rate of the potential is not simply
exponential, or the exponential decay rate is too fast, i.e.
$\beta>\sqrt{4 \mu_0}$, then the simple $c_0$ formula (\ref{e:c0})
in the previous subsection will be invalid. For instance, when
$\beta>\sqrt{4 \mu_0}$, the integral in the numerator of
(\ref{e:c0}) would be divergent in view of the asymptotics of its
integrand. The quantity on the right side of Eq. (75) would not
vanish either. Thus the solvability condition for these more general
potentials needs a different treatment.

In our new treatment, we consider the $R_1(x)$ solution (\ref{f:R1})
and demand that its tail asymptotics match (\ref{e:R1asym}). In
particular, this entails choosing $c_0$ such that the terms of which
decay like $g(x)$ ($e^{-\beta|x|}$ for exponential potentials) are
eliminated.

Suppose the tail asymptotics of the second homogenous solution
$\phi_2(x)$ has the form
\[  \label{e:phi2tail2}
\phi_2 \rightarrow \left\{ \begin{array}{c c} \tau^{-}(x), & x \rightarrow -\infty,
\\ \kappa_2 e^{-\sqrt{4\mu_0}x} + \tau^{+}(x), & x \rightarrow +\infty, \end{array} \right.
\]
where $\kappa_2$ is a certain real constant, and $\tau^{\pm}(x)$ are
the other decaying tail functions. This $\phi_2$ asymptotics is the
counterpart of Eqs. (\ref{e:asymphi2l})-(\ref{e:asymphi2r}) in the
previous subsection.

We substitute the $F$ formula (\ref{e:Fformula}) into
(\ref{f:R1}). Then this $R_1$ solution can be rewritten as
\[ \label{f:R1b}
R_1(x)=c_0\left[R_{11}(x)-c_0\sigma R_{12}(x)\right]+c_1\phi(x)+c_2\phi_2(x),
\]
where $R_{11}(x)$ and $R_{12}(x)$ are particular solutions of
equations
\[
L R_{11}(x)=f_1,\quad  L R_{12}(x)=f_2.
\]
For definiteness, we impose zero initial conditions on $R_{11}$ and
$R_{12}$ at $x=0$, i.e.,
\[  \label{e:R1kx0}
R_{1k}(0)=R'_{1k}(0)=0,  \quad k=1, 2.
\]
Notice that both particular solutions $R_{11}$ and $R_{12}$ approach
zero at large $x$, since the forcing terms $f_1$ and $f_2$ approach
zero, and the homogeneous solutions are all localized.

The tails of the particular solutions $R_{11} $ and $R_{12}$ each
have terms that decay exponentially and a term which decays like
$\tau^{\pm}(x)$, due to the exponentially decaying forcing terms and
exponential tails inside the homogeneous solutions. Specifically,
\begin{subequations} \label{e:R11R12asym}
\begin{eqnarray}
R_{11}(x) &\rightarrow &\left\{ \begin{array}{c c} P_{11}^{-}(x) e^{\sqrt{4\mu_0} x} +  \chi_{1}^- \tau^{-}(x), & x \rightarrow -\infty, \\
P_{11}^{+}(x) e^{-\sqrt{4\mu_0} x} +  \chi_{1}^+ \tau^{+}(x), & x \rightarrow +\infty, \end{array}\right.        \\
R_{12}(x) &\rightarrow& \left\{ \begin{array}{c c} P_{12}^{-} \hspace{0.05cm} e^{\sqrt{4\mu_0} x} +C_1^-e^{2\sqrt{4\mu_0} x}+ \chi_{2}^- \tau^{-}(x), & x \rightarrow -\infty, \\
P_{12}^{+} \hspace{0.05cm} e^{-\sqrt{4\mu_0} x} +C_1^+e^{-2\sqrt{4\mu_0} x}+  \chi_{2}^+ \tau^{+}(x), & x \rightarrow +\infty.  \end{array} \right.
\end{eqnarray}
\end{subequations}
Here $P_{11}^{\pm}$ are linear functions and $P_{12}^{\pm}$,
$C_1^{\pm}$ are constants.

Substituting these asymptotics into the $R_1(x)$ formula
(\ref{f:R1b}) and comparing its tails with Eq. (\ref{e:R1asym}), we
see that the coefficients on $\tau^{\pm}(x)$ must be zero as
$x\rightarrow \pm \infty$. This leads to the following system of
equations
\[  \label{e:c0chi}
c_0\chi_1^- -c_0^2\sigma \chi_2^{-}+c_2=0, \quad c_0\chi_1^+ -c_0^2\sigma \chi_2^{+}+c_2=0.
\]
From these, we obtain the necessary and sufficient solvability condition as
\[  \label{e:c0bb}
c_0= \frac{\chi_1^+ - \chi_1^- }{\sigma\left( \chi_2^+ - \chi_2^- \right)}.
\]

For potentials with exponential decay rates, the constants
$\chi_1^\pm, \chi_2^\pm$ can be found analytically with a bit of
effort. However, in general these constants in the tails of
$R_{11}(x)$ and $R_{12}(x)$ may not be known analytically, since the
tail behaviors $\tau^{\pm}(x)$ of the second homogeneous solution
$\phi_2(x)$ may not be analytically available. Regardless, these
constants $\chi_1^\pm$ and $\chi_2^\pm$ can be efficiently evaluated
numerically, as examples in the next subsection will show.

\subsection{Numerical examples} Now, we numerically confirm the above
analysis with two examples.

\textbf{Example 1}\, For the first example, we choose the complex
potential (\ref{e:Vx}) with an uneven double-hump function
\[ \label{e:gx}
g(x)=0.8\left[\sech(x+2)+h\, \sech(x-2)\right],
\]
where $h$ is a positive constant, and $\sigma=1$ (focusing
nonlinearity). Notice that this potential is exponentially decaying,
satisfying our Assumption 2 with $\beta=1$.

\begin{figure}[!htbp]
    \centering
    \includegraphics[width=1.0\textwidth]{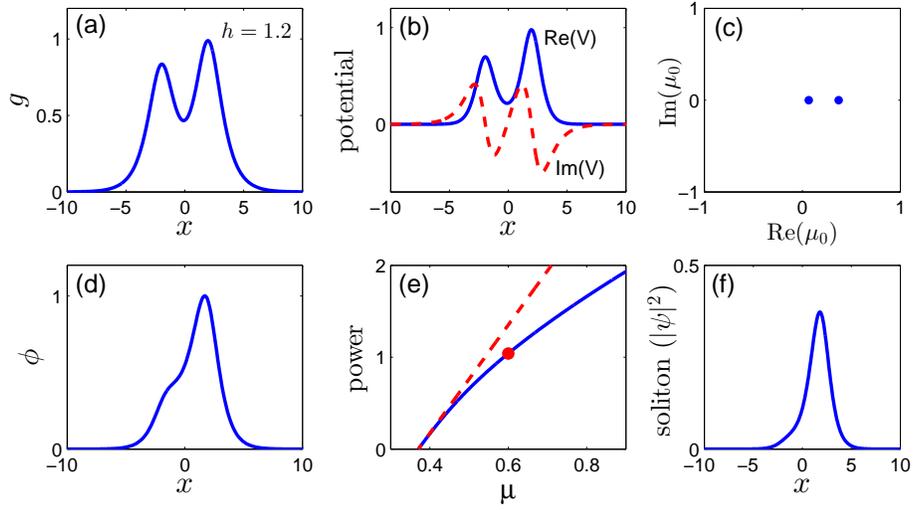}
    \caption{(a) The example function $g(x)$ in Eq. (\ref{e:gx}) with $h=1.2$;
    (b) Complex potential $V(x)$ for the $g(x)$ function in (a); (c) Eigenvalues $\mu_0$ of the eigenmode problem (\ref{e:phi});
(d) Eigenfunction $\phi$ of the largest eigenvalue in (c); (e) Power curve
    of solitons bifurcating from the linear mode in (d) under focusing nonlinearity
    [solid blue: numerical values; dashed red: the line with the theoretical slope value (\ref{e:slope})];
    (f) Soliton ($|\psi|^2$) at the marked point of the power curve.  }  \label{f:fig1}
\end{figure}

When $h=1.2$, the function $g(x)$ and the corresponding complex
potential $V(x)$ are displayed in Fig.~\ref{f:fig1}(a,b)
respectively. Notice that this potential is non-\PT-symmetric.
Eigenvalues of the eigenmode problem (\ref{e:phi}) are shown in
panel (c), where two real eigenvalues are found. The larger of these
eigenvalues is $\mu_0\approx 0.3708$, whose eigenfunction $\phi(x)$
is plotted in panel (d). For this eigenvalue, $\sqrt{4\mu_0}>1$,
thus Assumption 3 is met, and the analysis in
Sec.~\ref{subsec:solvability} applies.

From this linear eigenmode, we have verified numerically that a
continuous family of solitons bifurcates out. The power curve of
this soliton family is shown in panel (e). Here the power is defined
as $P=\int_{-\infty}^\infty |\psi|^2 dx$. The analytical prediction
for the power slope $P'(\mu_0)$ at the bifurcation point can be
obtained from equations (\ref{e:Rseries})-(\ref{e:R0}) as
\[ \label{e:slope}
P_{\rm anal}'(\mu_0)=c_0 \int_{-\infty}^\infty \phi dx,
\]
where $c_0$ is given by formula (\ref{e:c0}). For $h=1.2$, this
analytical power slope is found to be approximately 5.8961. The line
with this power slope is plotted as dashed red line in panel (e),
and good agreement with the numerical power slope can be seen. In
panel (f), the amplitude profile $R=|\psi|^2$ of the soliton at the
marked point of the power curve (with $\mu=0.6$) is displayed.

\begin{figure}[!htbp]
    \centering
    \includegraphics[width=1.0\textwidth]{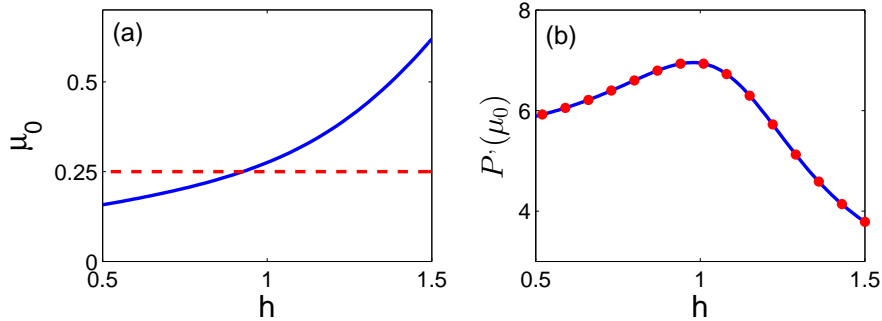}
    \caption{(a) Linear eigenvalue $\mu_0$ in Eq. (\ref{e:phi}) for the $g(x)$ function (\ref{e:gx}) at various $h$ values;
    the horizontal dashed line is at $\mu_0=1/4$;
    (b) comparison on the slope of the power curve at the bifurcation point $\mu_0$ between numerical values (solid blue) and analytical predictions
    (red dots).  }  \label{f:fig2}
\end{figure}

As parameter $h$ in the $g(x)$ function (\ref{e:gx}) varies, the
discrete eigenvalue $\mu_0$ will change [see Fig.~\ref{f:fig2}(a)].
When $h$ drops below 0.926, $\mu_0$ will fall under 0.25, entering
the $\sqrt{4\mu_0} < \beta$ regime (where Assumption 3 does not
hold). In order to test our theory for both $\sqrt{4\mu_0} > \beta$
and $\sqrt{4\mu_0} < \beta$ cases, we have plotted in
Fig.~\ref{f:fig2}(b) the theoretical predictions for the power slope
$P_{\rm anal}'(\mu_0)$ in Eq. (\ref{e:slope}) for $0.5\le h\le 1.5$,
which encompasses both cases. The reader is reminded that the $c_0$
formula is given by Eq. (\ref{e:c0}) when $\sqrt{4\mu_0} > \beta$
and by Eq. (\ref{e:c0bb}) when $\sqrt{4\mu_0} < \beta$. In the same
figure, numerically obtained power slopes $P'(\mu_0)$ for each $h$
value are shown as well. It is seen that numerical and analytical
slope values exactly match each other, confirming the accuracy of
our theoretical analysis in sections
\ref{subsec:solvability}-\ref{subsec:Generalization}.

\begin{figure}[!htbp]
    \centering
    \includegraphics[width=1.0\textwidth]{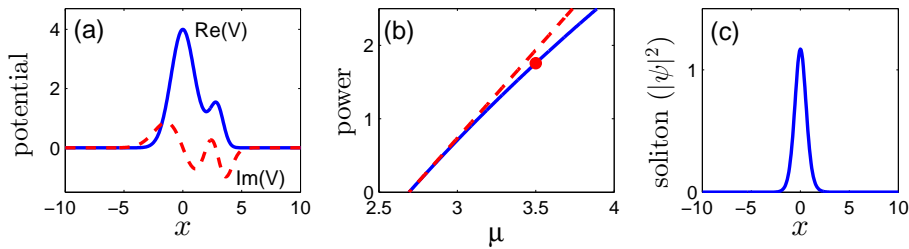}
    \caption{(a) Complex potential $V(x)$ for the $g(x)$ function in Example 2; (b)
    Power curve of solitons under focusing nonlinearity
    [solid blue: numerical values; dashed red: the line with the theoretical slope (\ref{e:slope})];
    (c) Soliton ($|\psi|^2$) at the marked point of the power curve.}  \label{f:fig3}
\end{figure}

\textbf{Example 2}\, As for the second example, we consider the
potential (\ref{e:Vx}) with
\[
g(x) = 2 e^{-x^2/4}+ e^{-(x-3)^2}.
\]
The resulting potential $V(x)$ is displayed in Fig.~\ref{f:fig3}(a).
The tails of this potential decay like a Guassian, which is faster
than exponential. Thus the results in
Sec.~\ref{subsec:Generalization} apply. In this case, analytical
expressions for the tail functions $\tau^{\pm}(x)$ of $\phi_2(x)$ in
Eq. (\ref{e:phi2tail2}) are not easy to obtain, but numerical approximations can be
readily computed. Specifically we select the $\phi_2(x)$ function by
requiring that for $x\rightarrow -\infty$ the function decay like a
gaussian and take this tail to be $\tau^-(x)$. Now for $x
\rightarrow +\infty$, the dominant decay of this tail is
exponential, i.e., the tail term $\tau^{+}(x)$ decays faster than
$e^{-\sqrt{4\mu_0} x}$ in Eq. (\ref{e:phi2tail2}), thus one must first
find the coefficient $\kappa_2$ of the exponential tail from
large-$x$ values of $\phi_2(x)$. Then subtracting away this
exponential tail from $\phi_2(x)$, the remaining tail is then
$\tau^+(x)$. To obtain $\chi_1^\pm$ and $\chi_2^\pm$ in Eq.
(\ref{e:R11R12asym}), we first compute $R_{11}(x)$ and $R_{12}(x)$
from the inhomogeneous equation (\ref{e:R1}), with $F$ replaced by
$f_1$ and $f_2$, under the initial conditions (\ref{e:R1kx0}). This
is done by integrating the inhomogeneous equation from $x=0$ out to
$x=\pm \infty$. By substracting their (slower-decaying) exponential
tails and comparing the remaining tails with $\tau^\pm(x)$ in
$\phi_2$, $\chi_1^\pm$ and $\chi_2^\pm$ can then be ascertained.
From these numbers, the $c_0$ value is calculated from formula
(\ref{e:c0bb}).

Now we compare these analytical predictions against numerical results.
Solving the eigenvalue problem (\ref{e:phi}), we find three discrete
real eigenvalues, the largest being $\mu_0\approx 2.6923$. From this
eigenmode, we have confirmed that a soliton family indeed bifurcates
out. If the nonlinearity is focusing ($\sigma=1$), the power curve
of this soliton family is plotted in Fig.~\ref{f:fig3}(b), and the
profile of the soliton at the marked point of the power curve (with
$\mu=3.5$) is illustrated in panel (c). On the power curve, the line
with analytically predicted power slope at the bifurcation point
from Eqs. (\ref{e:c0bb}) and (\ref{e:slope}) is also plotted. It is
seen that this analytical power slope matches the numerical one very
well.

\section{Extension to more general nonlinearities}

In this section, we show that the results in the previous sections
can be readily extended to a wider class of nonlinearities
\[  \label{e:NLSG}
i\Psi_t+\Psi_{xx}+V(x)\Psi+G(|\Psi|^2)\Psi=0,
\]
where $G(\cdot)$ is an arbitrary real function, and $V(x)$ is a
complex potential. Solitons (\ref{e:soliton}) in this equation
satisfy the stationary equation
\[ \label{e:psi2}
\psi_{xx}-\mu\psi+V(x)\psi+G(|\psi|^2)\psi=0.
\]

Just as in the case of cubic nonlinearity, in the absence of the
potential [$V(x)=0$], this soliton equation admits two constants of
motion. Assuming that the constant of motion in the presence of the
complex potential $V(x)$ is a continuous deformation of those
without the potential, we can show by the same technique employed in
Sec.~\ref{sec:3} that the only complex potentials which admit a
constant of motion are those in the special form of (\ref{e:Vx}),
and the corresponding constant of motion is
\[  \label{e:Cmotion2}
J=r_x^2-\mu r^2 +H(r^2)+r^2(\theta+g)^2,
\]
where $H(z)=\int G(z)dz$, and $dJ/dx=0$.

We can also show that for these general nonlinearities, with
potentials of the form (\ref{e:Vx}), continuous families of solitons still
bifurcate out from linear discrete eigenmodes. Without loss of
generality, we require $H(0)=0$. Then for solitons, $J=0$. Using
this relation, the equation for the complex soliton $\psi(x)$ is
reduced to the following second-order equation for the real
amplitude variable $R(x)=|\psi(x)|^2$:
\[  \label{e:Rpm2}
R_{xx}-4\mu R+2R\hspace{0.05cm} G(R)+2H(R)
\pm 2g \sqrt{4\mu R^2-4R\hspace{0.05cm} H(R)-R_x^2}=0.
\]
This equation is the analog of Eq. (\ref{e:Rpm}) for the cubic NLS
equation (\ref{e:NLS}). Repeating similar analysis as in the earlier
text, these soliton bifurcations can be explicitly calculated.

To illustrate these analytical results for general nonlinearities,
we consider the following example with a saturable nonlinearity.

\begin{figure}[!htbp]
    \centering
    \includegraphics[width=1.0\textwidth]{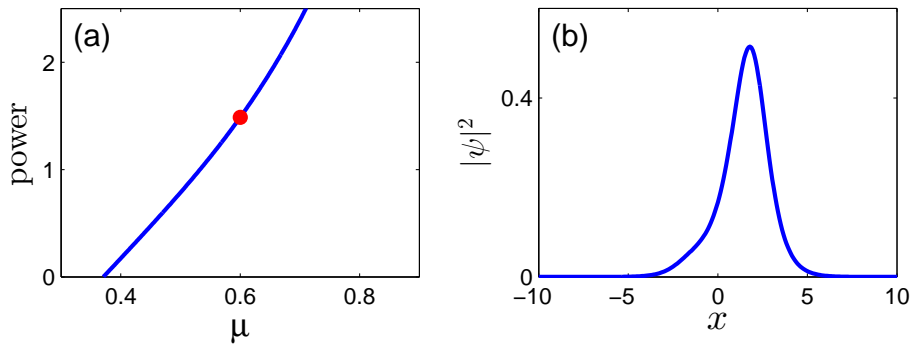}
    \caption{(a) Power curve
    of solitons bifurcating from a linear mode in Example~3;
    (b) Soliton ($|\psi|^2$) at the marked point of the power curve.    }  \label{f:fig4}
\end{figure}

\textbf{Example 3}\, Consider the NLS equation (\ref{e:NLSG}) with a
saturable nonlinearity and complex potential,
\[  \label{e:NLSGE}
i\Psi_t+\Psi_{xx}+V(x)\Psi+\frac{|\Psi|^2}{1+|\Psi|^2}\Psi=0,
\]
where the potential $V(x)$ is of the special form (\ref{e:Vx}) with
$g(x)$ chosen the same as in Example 1 [i.e., $g(x)$ is given
by Eq. (\ref{e:gx})], with $h$ fixed as $h=1.2$. Solitons in this
equation are sought of the form (\ref{e:soliton}), where $\psi(x)$
solves
\[ \label{e:psi11}
\psi_{xx}-\mu\psi+V(x)\psi+\frac{|\psi|^2}{1+|\psi|^2}\psi=0.
\]
Since the potential here is the same as that in Example 1, discrete
eigenvalues in the linear equation (\ref{e:phi}) remain the same as
those shown in Fig.~\ref{f:fig1}(c), with the larger one being
$\mu_0\approx 0.3708$. From this eigenmode, we have confirmed that a
continuous family of solitons bifurcates out, whose power curve is
displayed in Fig.~\ref{f:fig4}(a). At the marked point of the power
curve, the corresponding soliton is plotted in Fig.~\ref{f:fig4}(b).
This example verifies that the bifurcation of soliton families in
complex potentials (\ref{e:Vx}) occurs for a wider class of
nonlinearities (\ref{e:NLSG}).

\section{Summary and discussion}

In this paper, we have analyzed soliton families in NLS-type
equations with non-\PT-symmetric complex potentials. Under a weak
assumption, we have shown that stationary forms of these equations
admit a constant of motion if and only if the complex potential is
of the special form (\ref{e:Vx}). Using this constant of motion, we
reduced the second-order complex soliton equation to a new
second-order real equation for the amplitude of the soliton. From
this new soliton equation, we showed, by perturbation methods, that
continuous families of solitons always bifurcate out from linear
eigenmodes for this special form of complex potentials. These
results hold not only for the cubic nonlinearity, but also for a
much wider class of nonlinearities. While it has been known that
\PT-symmetric dissipative systems share some important properties with
conservative systems, the results in this paper reveal that certain
types of non-\PT-symmetric dissipative systems can also share such
properties of conservative systems (such as the existence of soliton
families).

Our results also shed light on a more general question: what
non-\PT-symmetric complex potentials in the NLS-type equations
(\ref{e:NLS}) and (\ref{e:NLSG}) admit continuous families of
solitons? In the absence of \PT symmetry, the existence of a
constant of motion in the stationary soliton equation is critical
for the existence of soliton families. We have shown that such a
constant of motion exists only for special potentials of the form
(\ref{e:Vx}), assuming this constant of motion is a continuous deformation
of that from the potential-free equation. Since this assumption is
reasonable, we conjecture that \emph{the only non-\PT-symmetric
complex potentials which admit soliton families are those of the
special form (\ref{e:Vx}).}

It should be pointed out that the question of solitons in
non-\PT-symmetric potentials is closely related to the question of
non-\PT-symmetric solitons in \PT-symmetric potentials. Indeed, for
\PT-symmetric potentials of the same special form (\ref{e:Vx}),
where $g(x)$ is taken to be even, it has been shown numerically that
symmetry breaking of solitons can occur \cite{Yang_OL2004}. As a
consequence, continuous families of non-\PT-symmetric solitons exist
in a \PT-symmetric potential. This symmetry breaking is 
surprising since it is forbidden in generic \PT-symmetric potentials
\cite{Yang_SAPM2014}. Analytical understanding of this symmetry
breaking is still an open question, however, based on the analysis in this
paper, it is hopeful that this symmetry breaking can now be
analytically studied. But this lies outside the scope of the present
article.

In the end, we mention that bifurcation of soliton families from
linear modes occurs in special forms of two-dimensional
non-\PT-symmetric complex potentials as well \cite{Yang_PRE2015}.
Analytical understanding of such bifurcations in two spatial
dimensions is a more challenging question which merits further
investigation.

\section*{Acknowledgment} This work was supported in part by the Air Force Office of
Scientific Research (USAF 9550-12-1-0244) and the National Science
Foundation (DMS-1311730).

\end{document}